\documentclass[12pt,a4]{article}
\usepackage{graphicx}
\usepackage{amsmath,amssymb}

\newcommand{\be}{\begin{equation}}
\newcommand{\ee}{\end{equation}}
\newcommand{\bea}{\begin{eqnarray}}
\newcommand{\eea}{\end{eqnarray}}
\newcommand{\bean}{\begin{eqnarray*}}
\newcommand{\eean}{\end{eqnarray*}}
\newcommand{\beq}{\begin{eqnarray}}
\newcommand{\eeq}{\end{eqnarray}}

\def\Na{^{22} \mathrm{Na}}

\relax

\begin{document}

\begin{center}

{\Huge Maximal angular correlation in $\gamma-\gamma$ coincidences: a quantitative study}

\vskip 10mm

Filipe Moura

\vskip 4mm
Departamento de Matem\'atica, Escola de Tecnologias e Arquitetura, \\ ISCTE - Instituto Universit\'ario de Lisboa \\ and Instituto de Telecomunica\c c\~oes,
\\Av. das For\c cas Armadas, 1649-026 Lisboa, Portugal
\vskip 4mm

{\tt fmoura@lx.it.pt}

\vskip 6mm

{\bf Abstract } \end{center}
\begin{quotation}\noindent
The measurement of the angular distribution of maximally correlated annihilation gamma rays radiated in coincidence, like those emitted from a $\Na$ source, is a classic experiment that is nowadays ordinarily performed in Nuclear Physics laboratory classes. For the first time we present an analytic expression for such angular distribution, which can be easily tested and confronted with the laboratory measurements.
\end{quotation}

\vfill
\eject
\newpage
%\tableofcontents
%%%%%%%%%%%%%%%%%%%%%%%%%%%%%%%%%%%%%%%%%%%%%%%%%%%%%%%%%%%%%%%%%%%%%%
%%%%%%%%%%%%%%%%%%%%%%%%%%%%%%%%%%%%%%%%%%%%%%%%%%%%%%%%%%%%%%%%%%%%%%

%%%%%%%%%%%%%%%%%%%%%%%%%%%%%%%%%%%%%%%%%%%%%%%%%%%%%%%%%%%%%%%%%
%%%%%%%%%%%%%%%%%%%%%%%%%%%%%%%%%%%%%%%%%%%%%%%%%%%%%%%%%%%%%%%%%
\section{Introduction}
%%%%%%%%%%%%%%%%%%%%%%%%%%%%%%%%%%%%%%%%%%%%%%%%%%%%%%%%%%%%%%%%%
%%%%%%%%%%%%%%%%%%%%%%%%%%%%%%%%%%%%%%%%%%%%%%%%%%%%%%%%%%%%%%%%%
\label{int}
\indent

Angular correlations of gamma rays emitted in coincidence by a radioactive source have been studied for a long time \cite{h40,y48,bd50,Biedenharn:1953yw}. In order to find such correlations, one measures the variation of the counts of coincidences, by keeping one of the detectors fixed and rotating another one around the radioactive source. The angular correlation function depends only on the relative angle between the emitted photons, while the experimental distribution of the counts of coincidences should clearly be an explicit function of the angle of rotation of the detector around the source.

If the spatial resolution of the detectors were perfect (i.e. if one could perfectly reconstruct the trajectory of a detected photon), two photons in coincidence would be detected only if their relative angle of emission was precisely equal to the angle of rotation of the detector. If that was the case, at a given angle the count of coincidences would simply be proportional to the angular correlation function. But that is not the case, and one must therefore account for the effects of the finite size of the detectors. Because of such effects, the relative angle of the detectors (having the source as the vertex), when detecting a coincidence, is not necessarily equal to the relative angle of emission of the respective pair of photons. Indeed, for a given angle between the photons there is a finite range of angles between the detectors such that the coincidence of the photon pair can be detected. This is why the experimental angular distribution is a smeared version of the theoretical correlation function \cite{lf53,rsc}. Such smearing is expressed in terms of geometrical corrections given by coefficients related to the finite solid angle subtended by each detector \cite{af55}. But because of the relation between the ``photon'' angle and the ``detector'' angle, the geometry of the problem at each concrete case is often complicated. That is the reason why one may find in the literature theoretical expressions for the angular correlation functions of different sources \cite{h40,bd50}, but not for the experimental angular distributions. This is because the coefficients expressing the finite solid angle corrections are not determined from the geometry of the problem (as they could be in principle), but always rather left as parameters to be determined by fitting the experimentally measured counts of coincidences \cite{cruz,barrette}.

The study of angular correlations and coincidence counts of fully correlated gamma rays, like those emitted from a $\Na$ source, is nowadays a standard classroom experiment in undergraduate Physics courses \cite{Melissinos,ortec,Ferguson}. In this case, the angular correlation function is trivial and the geometrical analysis is \emph{a priori} much simpler. Yet, an expression for the expected experimental angular distribution of the counts of coincidences does not exist even for this case.

The goal of this article is to fill that gap in the literature and provide an analytical expression for such distribution in terms of variables that can be determined in the laboratory, so that a more accurate comparison between the predicted and the measured counts can be made.

The article is organized as follows. In section \ref{setup} we review the main concepts associated with this study. We distinguish between uncorrelated and correlated pairs of gamma rays, we compute the geometrical efficiency associated with a detector and, based on that efficiency, we determine the rate of coincidences for pairs of uncorrelated gamma rays. In section \ref{calc}, we compute the rate of coincidences for pairs of fully correlated gamma rays. Finally, in section \ref{conc} we discuss our results and further possible extensions.

%%%%%%%%%%%%%%%%%%%%%%%%%%%%%%%%%%%%%%%%%%%%%%%%%%%%%%%%%%%%%%%%%
%%%%%%%%%%%%%%%%%%%%%%%%%%%%%%%%%%%%%%%%%%%%%%%%%%%%%%%%%%%%%%%%%
\section{Setup and description of the experiment}
%%%%%%%%%%%%%%%%%%%%%%%%%%%%%%%%%%%%%%%%%%%%%%%%%%%%%%%%%%%%%%%%%
%%%%%%%%%%%%%%%%%%%%%%%%%%%%%%%%%%%%%%%%%%%%%%%%%%%%%%%%%%%%%%%%%
\label{setup}
\indent

Consider the detection of an emission of specific gamma rays by some radioactive nucleus. The
counting rate, in $\mathrm{s}^{-1}$, of the counter associated to the detector is given by
\be
N = A \epsilon \frac{\Omega}{4 \pi}. \label{count}
\ee
Here $A$ is the number of gamma rays per second emitted by the isotropic radioactive source: it is given as the product of the source activity times the fraction of its decays that result in the gamma rays we are considering. $\epsilon$ is the \emph{intrinsic efficiency} of the detector for the corresponding gamma ray energy, and $\Omega$ is the solid angle subtended at the source by the face of the detector. One often calls the fraction $\frac{\Omega}{4 \pi}$ the \emph{geometrical efficiency} of the detector \cite{Leo:1987kd}.

Indeed, we assume that the face of the detector is a circle of finite radius $r$, at a distance $d$ from the isotropic source it is facing. The fraction of events impinging on the detector over the events emitted by the source corresponds to the area of the spherical cap limited by the face of the detector (and inserted on a sphere having the source at its center) divided by the area of the whole sphere. The radius of that sphere is $R= \sqrt{d^2 + r^2}$, and its area is of course $4 \pi R^2$. Defining the angle
\be
\beta=\arctan \frac{r}{d},  \label{bet}
\ee
the area of the spherical cap is given by an elementary surface integral as $2 \pi R^2 \left(1-\frac{d}{R}\right)$ or $2 \pi R^2 \left(1-\cos \beta\right)$. Dividing by the area of the sphere $4 \pi R^2$, we get for the geometrical efficiency in (\ref{count}) \cite{Knoll:2000fj}
\be
\frac{\Omega}{4 \pi}=\frac{1}{2} \left(1-\cos \beta\right). \label{solid}
\ee
The value of $\beta$ can be varied by choosing the distance $d$ between the detector and the source (of course, given the restrictions in each laboratory). Typical values of $\beta$ are small. Smaller values of $\beta$ mean a larger angular resolution but a smaller geometrical efficiency.

Consider now the almost simultaneous emission of two gamma rays, each one being detected by its own detector. Depending on if their emissions are independent events or not, the respective photons are said to be \emph{uncorrelated} or, otherwise, \emph{correlated}. If their detections occur simultaneously, the photons are said to be in \emph{temporal coincidence}. The relative probability that a photon will be emitted at an angle $\theta$ with respect to a previously emitted photon is called the angular distribution function and denoted $W(\theta)$.

In general, when two gamma rays are emitted in succession from an atomic
nucleus, their directions are correlated due to the physics
of the emission process. In particular, when an excited nuclear
state decays to the ground state through one or more
intermediate states, the spin of the nucleus affects the angular
distribution of the photons emitted during each transition.
In these cases, the angular distribution function
$W(\theta)$ can depend both on the spin of the states
involved in the transitions and on the multipole order of the
emitted radiation \cite{h40}. In the limit case when the two photons are uncorrelated, the angular distribution function is uniform and isotropic: $W(\theta)= \frac{1}{2\pi}$.

The other limit case occurs in the simple process when an isotope undergoes $\beta^+$ decay, after which the resulting positron is captured by an electron, and they both annihilate to produce a pair of 511 keV gamma rays. Because of conservation of momentum, the two photons must be
emitted in exactly opposite directions, with a relative angle $\theta=\pi$. Therefore, in this case the two photons are \emph{totally correlated} and the corresponding angular distribution function is simply given by
\be
W(\theta) = \delta (\theta - \pi); \label{delta}
\ee
this is what happens with sodium-22 ($\Na$).

Studying angular correlations can be very useful for the analysis of nuclear decay schemes and the assignment of spin and parity to excited nuclear states. Assuming one of the detectors is fixed and the other one can be moved along a circumference having the source at its center, the rate of coincident counts (i.e. the counts of photons in temporal coincidence) can be measured for different values of the position of the moving detector, which can in principle be identified with the angle $\theta$. Up to an overall normalization, this rate can be identified with the angular distribution function, at least if the correlation between the pair of gamma rays being considered is not very high.

Let $ \epsilon_1, \, \epsilon_2$ be the intrinsic efficiencies and $\Omega_1, \, \Omega_2$ be the solid angles subtended by the faces of the two detectors associated to the coincidence counts. According to (\ref{count}), their counting rates are respectively given by
\bea
N_1 = A \epsilon_1 \frac{\Omega_1}{4 \pi}, \label{count1}\\
N_2 = A \epsilon_2 \frac{\Omega_2}{4 \pi}. \label{count2}
\eea

If the pair of gamma rays being considered is uncorrelated, or if its correlation is low, the rate of ``true'' coincidences is given by \cite{Melissinos,Knoll:2000fj} \footnote{Here we mean ``true'' coincidences, coming from events originating in the same decay. They should be distinguished from ``accidental'' coincidences, due to the accidental combination of two separate events from independent decays that occur closely spaced in time.}
\be
C_U = A \epsilon_1 \epsilon_2 \frac{\Omega_1}{4 \pi} \frac{\Omega_2}{4 \pi}. \label{coinu}
\ee
The solid angles $\Omega_1, \, \Omega_2$ can be obtained from (\ref{solid}), assuming for each detector geometric configurations with angles $\beta_1, \, \beta_2$. Like equation (\ref{count}), (\ref{coinu}) contains an overall normalization factor, $A$, and probabilistic factors, the intrinsic and geometrical efficiencies. This formula evidences the fact that, since the emissions of the gamma rays are independent events, their joint probability is the product of the probabilities of each separate event. It does not have any dependence on the angle $\theta$, as it should.

For correlated photons, there is a strong angular dependence on the rate of coincidence counts, as we mentioned. In general, the rate of coincidences includes a factor depending on the correlation of the pair of photons (for a discussion see \cite{Knoll:2000fj}); such correlation is quite difficult to determine in principle. But in the limit when the gamma rays are fully correlated, like when the angular distribution function is given by (\ref{delta}), the rate of coincidences can be worked out. That is the main goal of this article.

%%%%%%%%%%%%%%%%%%%%%%%%%%%%%%%%%%%%%%%%%%%%%%%%%%%%%%%%%%%%%%%%%
%%%%%%%%%%%%%%%%%%%%%%%%%%%%%%%%%%%%%%%%%%%%%%%%%%%%%%%%%%%%%%%%%
\section{Angular distribution of coincidence counts for maximally correlated gamma rays}
%%%%%%%%%%%%%%%%%%%%%%%%%%%%%%%%%%%%%%%%%%%%%%%%%%%%%%%%%%%%%%%%%
%%%%%%%%%%%%%%%%%%%%%%%%%%%%%%%%%%%%%%%%%%%%%%%%%%%%%%%%%%%%%%%%%
\label{calc}
\indent

\begin{figure}[h]
\centering
\includegraphics[width=100mm]{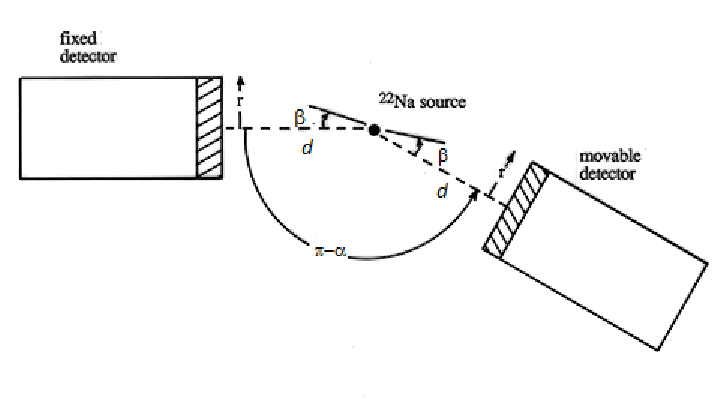}
\caption{\small{The fixed and the rotating detectors and definitions of $\alpha, \, \beta, \, r, d$.}}
\label{fig1}
\end{figure}

For the remainder of the article, we will assume the existence of two detectors. For simplicity, we will assume that the two detectors are identical (that is what typically happens in practice, although the following discussion can be generalized to detectors having different geometries). Yet, the detectors are distinguishable: one of them is fixed, and the other one can be moved along a circumference. In the center of that circumference, there is a source of $\Na$, emitting pairs of gamma rays in opposite directions, with an angular distribution function given by (\ref{delta}). Each of the two detectors is placed facing the $\Na$ source (fig. \ref{fig1}).

If the detectors were point-like, in order to detect a coincidence they would have to be also facing each other, in a straight line. The angular dependence of the rate of coincidences would be similar to (\ref{delta}) and, up to a normalization, it could be identified with the angular distribution function, as we mentioned. But the detectors have a finite size, expressed by the finite solid angle $\Omega$ in (\ref{count}). Because of that finite size, it is possible to detect coincident photons as long as the two detectors are placed so that they can be hit by them, even if the detectors are not facing each other. To illustrate that, let's make a small geometric digression.

As we saw, each of the detectors has a face, a circle of radius $r$ at a distance $d$ from the source. Associated with each detector there is a spherical cap, limited by its face, and corresponding to a solid angle $\Omega$. Consider now the reflection of the spherical cap of the moving detector across the plane that passes by the source and is perpendicular to the fixed detector. If the moving detector is in its original position, facing the fixed detector, opposite to it, the reflected spherical cap of the moving detector will coincide with the original spherical cap of the fixed detector. The counts in each of the detectors should be given by (\ref{count}), and they should match the number of coincidences, since each count would correspond to a coincidence.
But if one rotates the moving detector by some angle $\alpha$ around a circumference of radius $d$ (so that its distance to the source remains constant), the positions of the reflected and the original spherical caps no longer coincide. If the angle $\alpha$ of displacement of the moving detector is smaller than a critical value, the intersection of the two spherical caps is not empty. In this configuration, for each detector there are photons hitting it that do not reach the other detector. From (\ref{count}), the probability that a photon emitted from the source hits the detector is given by the geometrical efficiency $\frac{\Omega}{4 \pi}$. The probability that each of the two fully correlated photons hits a detector, therefore forming a coincidence, is therefore, by the same reasoning, given by
\be
C_C = A \epsilon_1 \epsilon_2 \frac{\Delta \Omega}{4 \pi}, \label{coinc}
\ee
$\Delta \Omega$ being the solid angle corresponding to the intersection of the original (fixed) and the reflected (rotating) spherical caps. $\frac{\Delta \Omega}{4 \pi}$ is the geometrical efficiency corresponding to the counting of coincidences. This is what we wish to compute, in terms of geometrical variables which can be determined in the laboratory.

\begin{figure}[h]
\centering
\includegraphics[width=100mm]{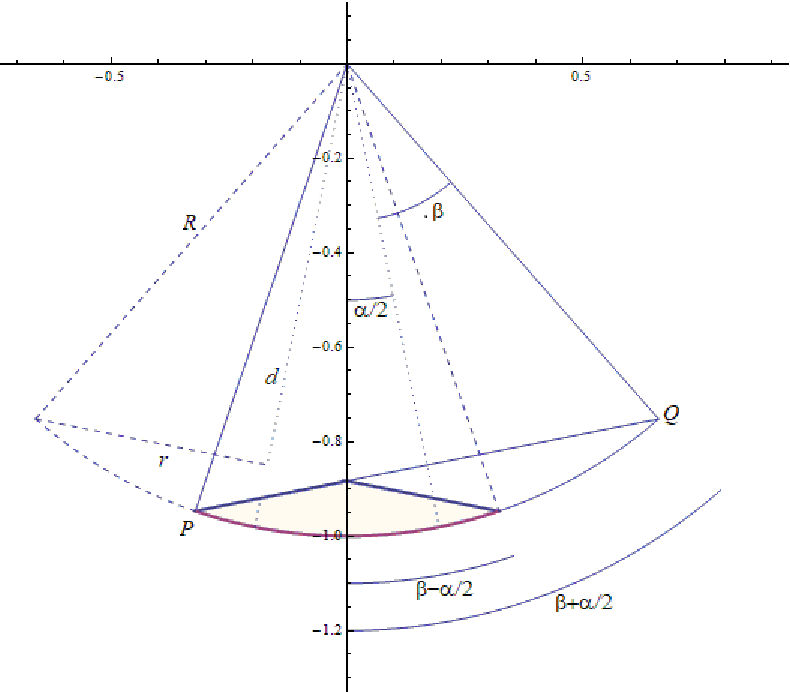}
\caption{\small{The fixed and the rotating spherical caps and their intersection. The original (fixed) cone is drawn in full line, while the reflected (rotating) cone is dashed.}}
\label{fig2}
\end{figure}

For values of $\alpha$ larger (modulo $2\pi$, of course) than a critical value, the intersection of the two caps becomes empty. That critical value of $\alpha$ can be easily determined from fig. \ref{fig2}; it is given by
\be
\alpha_{\mathrm{crit}}=2\beta. \label{crit}
\ee
$\Delta \Omega$ (and the number of coincidences) reach a maximum when $\alpha \equiv 0$ and a minimum (0) when $\alpha \equiv \alpha_{\mathrm{crit}}$, always modulo $2\pi$.

Despite the area in which we are interested being embedded into a sphere, the problem of calculating it does not have spherical symmetry. Indeed, such symmetry is broken by the existence of an axis around which one of the detectors rotates. Therefore spherical coordinates are not the most suitable for this case. But one can use normal cartesian coordinates. We take the plane of rotation as the $x-y$ plane, with the source located at the origin. The rotation takes place therefore around the $z$ axis. Initially the two detectors are facing each other. One then rotates the moving detector by an angle $\alpha$ and so does the spherical cap it defines. For convenience, we place the center of the face of the fixed detector at the position $(d \sin\left(\frac{\alpha}{2}\right), -d \cos\left(\frac{\alpha}{2}\right), 0)$, while the center of the one of the moving detector is at $(d \sin\left(\frac{\alpha}{2}\right), d \cos\left(\frac{\alpha}{2}\right), 0)$. The center of the reflected face of the moving detector is at $(-d \sin\left(\frac{\alpha}{2}\right), -d \cos\left(\frac{\alpha}{2}\right), 0)$. In fig. \ref{fig2}, we can see the projection in the $x-y$ plane of the intersection of the original fixed and of the reflected moving spherical caps. We want to compute the area of such intersection.

It is straightforward to figure out that points $P, Q$ in fig. \ref{fig2} have coordinates
\bean
P& \rightarrowtail &\left(-R \sin\left(\beta-\frac{\alpha}{2}\right), -R \cos\left(\beta-\frac{\alpha}{2}\right), 0\right), \\
Q& \rightarrowtail &\left(R \sin\left(\beta+\frac{\alpha}{2}\right), -R \cos\left(\beta+\frac{\alpha}{2}\right), 0\right),
\eean
from which we get the equation of the plane, perpendicular to the $x-y$ plane, which passes through these two points (represented by a straight line in fig. \ref{fig2}): $$y=\tan\left(\frac{\alpha}{2}\right)\left(x+R \sin\left(\beta-\frac{\alpha}{2}\right) \right)-R \cos\left(\beta-\frac{\alpha}{2}\right).$$
The surface whose area we wish to compute can be split into four equal parts. The first one can be seen as the surface plot of $Z(x, y)=\sqrt{R^2-x^2-y^2}$ (positive $z$ coordinate, above the $x-y$ plane), with $x-y$ range defined in the third quadrant ($x<0, y<0$) as
\bean
&& -\sqrt{R^2-x^2}\leq y \leq \tan\left(\frac{\alpha}{2}\right)\left(x+R \sin\left(\beta-\frac{\alpha}{2}\right) \right)-R \cos\left(\beta-\frac{\alpha}{2}\right), \\ &&-R \sin\left(\beta-\frac{\alpha}{2}\right) \leq x \leq 0.
\eean
The second part is given as the surface plot of the same function, but with $x-y$ range defined in the fourth quadrant ($x>0, y<0$). This range is analogous to the one of the first part, but reflected around the $y$ axis.
The third and fourth parts have the same $x-y$ ranges of the first and second part, respectively, but they are the surface plots of $Z(x, y)=-\sqrt{R^2-x^2-y^2}$ (negative $z$ coordinate, below the $x-y$ plane). They are the reflections of the first and second parts around the $x-y$ plane. Clearly the areas of the four parts are equal, and we can take for the total area four times the area of the first part, given by the following surface integral:
\be
4 \int_{-R \sin\left(\beta-\frac{\alpha}{2}\right)}^{0} \int_{-\sqrt{R^2-x^2}}^{\tan\left(\frac{\alpha}{2}\right)\left(x+R \sin\left(\beta-\frac{\alpha}{2}\right) \right)-R \cos\left(\beta-\frac{\alpha}{2}\right)} \frac{R}{\sqrt{R^2-x^2-y^2}} \, dy \, dx
\ee
Because we are only interested in the solid angle, we can simply take $R=1$ in the previous formula, obtaining:
\be
\Delta \Omega=4 \int_{- \sin\left(\beta-\frac{\alpha}{2}\right)}^{0} \int_{-\sqrt{1-x^2}}^{\tan\left(\frac{\alpha}{2}\right)\left(x+ \sin\left(\beta-\frac{\alpha}{2}\right) \right) - \cos\left(\beta-\frac{\alpha}{2}\right)} \frac{1}{\sqrt{1-x^2-y^2}} \, dy \, dx
\ee
After performing the integrations, the final result is given by
\be
\Delta \Omega \left(\alpha, \beta\right) =4 \left(\mathrm{arccot}\left(\frac{\sqrt{2} \sin \frac{\left|\alpha\right|}{2}}{\sqrt{\cos \alpha - \cos 2\beta}}\right) - \mathrm{arccot}\left(\frac{\sqrt{2} \cos \beta \sin \frac{\left|\alpha\right|}{2}}{\sqrt{\cos \alpha - \cos 2\beta}}\right)  \cos \beta \right), \label{omega}
\ee
with $\mathrm{arccot} (x) = \frac{\pi}{2} - \arctan(x)$ being the inverse cotangent function and $\left|\alpha\right|$ the absolute value of the angle $\alpha$.

We can check that the result (\ref{omega}) for $\Delta \Omega \left(\alpha, \beta\right)$ has some properties that one should expect. For a given value of $\beta$, it is a periodic function of $\alpha$, with period $2\pi$, for the values of $\alpha$ where it is defined. On a neighborhood of $\alpha=0$, it is defined (and positive) only for $\alpha \leq 2 \beta$. It vanishes for $\alpha \equiv 2 \beta$, the critical value $\alpha_{\mathrm{crit}}$ from (\ref{crit}). It is not defined as a real function in the intervals $2\beta < \alpha < 2\pi -2\beta$ and $-2\pi+2\beta < \alpha < -2\beta$, but that does not have any physical or geometrical meaning: for those values of $\alpha$, the rate of coincidences should be 0. For $\alpha=0$ (no rotation of the moving detector), (\ref{omega}) reduces to (\ref{solid}), as it should. In the limit $\beta \rightarrow 0$, when the detectors become point-like, (\ref{omega}) reduces to (\ref{delta}) with $\alpha=\pi-\theta$: in this limit, the angular resolution of the detectors is the highest, and the rate of coincidences matches the angular distribution function for the given source, as we saw.

\begin{figure}[h]
\centering
\includegraphics[width=100mm]{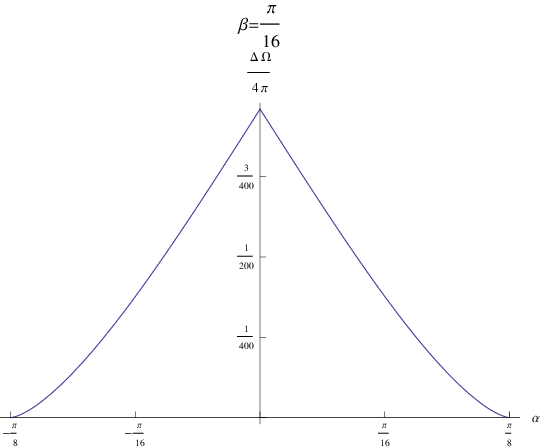}
\caption{\small{Plot of $\Delta \Omega/4\pi$ as a function of the rotation angle $\alpha$ for $\beta=\frac{\pi}{16}$.}}
\label{fig3}
\end{figure}

In fig. \ref{fig3} we present a plot of the fraction $\frac{\Delta \Omega}{4\pi}$ as a function of $\alpha$ for $\beta=\frac{\pi}{16}$, a value which may be considered reasonable for a typical laboratory configuration and similar to the one corresponding to the experiment described in \cite{ortec}.
From the plot we can see that, according to (\ref{crit}), $\alpha_{\mathrm{crit}}=\frac{\pi}{8}.$ The maximum value of $\frac{\Delta \Omega}{4\pi}$ is indeed obtained for $\alpha=0$ and given by $\frac{1}{2} \left(1-\cos \frac{\pi}{16}\right) \approx 0.0096,$ according to (\ref{solid}). The shape of this plot is similar to the ones of experimental plots of the rate of coincidences as a function of the rotation angle (usually expressed in degrees), which can be found for instance in \cite{bd50,ortec} and in the book \cite{Melissinos}. Only the values in the $y$ axis are very different, but here we are just plotting $\frac{\Delta \Omega}{4\pi}$, while in the experimental plots in the $y$ axis we find the rate of coincidences, related through (\ref{coinc}) by an overall factor $A \epsilon_1 \epsilon_2$ depending on the source activity and the detector intrinsic efficiencies.

The angle $\beta$ can be directly estimated in the laboratory from quantities which can be measured, using (\ref{bet}). Alternatively, one can also determine $\beta$ from (\ref{crit}) as half of the critical rotation angle $\alpha_{\mathrm{crit}}$ for which the rate of coincidences vanishes; such angle can also be directly estimated from experimental data. The gamma ray source activity $A$ and the detector intrinsic efficiencies $\epsilon_1, \, \epsilon_2$ can also be obtained experimentally; indeed, considering the previously described experimental configuration, the rates of gamma rays in each detector $N_1, \, N_2$ are given respectively by (\ref{count1}) and (\ref{count2}), with $\Omega_1=\Omega_2=\Omega$ given by (\ref{solid}). Without any detector rotation, i.e. with $\alpha=0$, the rate of gamma ray coincidences $C_C$ is given by (\ref{coinc}), with $\Delta \Omega=\Omega$ also given by (\ref{solid}). Solving these three equations (\ref{count1}), (\ref{count2}) and (\ref{coinc}) for $A, \, \epsilon_1, \, \epsilon_2$, we see that these three quantities can be expressed in terms of other quantities that are directly measurable in the laboratory \cite{ortec2}:
\bea
A&=&\frac{2}{1-\cos \beta} \frac{N_1 \, N_2}{C_C \left(\alpha=0\right)}, \label{cl}\\
\epsilon_1&=&\frac{C_C \left(\alpha=0\right)}{N_2}, \label{e1l} \\
\epsilon_2&=&\frac{C_C \left(\alpha=0\right)}{N_1}. \label{e2l}
\eea
The experimental results for the rate of coincidences as a function of the rotation angle $\alpha$ can be fitted to (\ref{coinc}) using the expression (\ref{omega}) we have derived, taking as fitting parameters the overall factor $A \epsilon_1 \epsilon_2$ and the angle $\beta$. The obtained fitted results should be compared to the ones directly and independently obtained in the laboratory, from (\ref{bet}) (or (\ref{crit})), (\ref{cl}), (\ref{e1l}), (\ref{e2l}), as a consistency test.

In the book \cite{Melissinos} there is also a detailed presentation of the experiment of measuring the angular correlations in $\gamma-\gamma$ coincidences, focusing on the $\Na$ source, including a complete description of the experimental apparatus. One can also find there estimates of the rate of accidental coincidences and of coincidences due to other uncorrelated gamma rays emitted by the same source (the 1.277 MeV gamma ray from the $^{22} \mathrm{Ne}$ neon decay), showing that they are both neglectable when compared to the rate of coincidences we have been considering.

\newpage
%%%%%%%%%%%%%%%%%%%%%%%%%%%%%%%%%%%%%%%%%%%%%%%%%%%%%%%%%%%%%%%%%
%%%%%%%%%%%%%%%%%%%%%%%%%%%%%%%%%%%%%%%%%%%%%%%%%%%%%%%%%%%%%%%%%
\section{Conclusions}
%%%%%%%%%%%%%%%%%%%%%%%%%%%%%%%%%%%%%%%%%%%%%%%%%%%%%%%%%%%%%%%%%
%%%%%%%%%%%%%%%%%%%%%%%%%%%%%%%%%%%%%%%%%%%%%%%%%%%%%%%%%%%%%%%%%
\label{conc}
\indent

In this article, we have derived an expression (\ref{omega}) for the geometrical factor which, together with (\ref{coinc}), gives us the angular distribution of the rate of coincidences for pairs of fully correlated gamma rays like those emitted by a $\Na$ source. Although the final result (\ref{omega}) seems simple, the calculation associated with it is quite nontrivial and was not yet available in the literature.

Previously, in order to determine such angular distribution one could just measure the rates of counts at different angles and check that there was a peak at the central angle. Having obtained (\ref{omega}), one can now actually go further and fit the theoretical rate of coincidences (\ref{coinc}) to experimental data, as a function of the rotation angle $\alpha$. From such fitting, one should reobtain the remaining parameters in (\ref{coinc}), corresponding to the angle $\beta$ related to the geometry of the detector and the overall factor $A \, \epsilon_1 \, \epsilon_2$, given by the product of the source activity and the intrinsic efficiencies of the detectors.

Here we should mention that the main purpose of this fitting which we propose, and of the experiment we have been considering, is not to directly measure any of the fitting parameters. As we have discussed, the physical quantities corresponding to the fitting parameters can be directly and independently obtained in the laboratory using much simpler and more accurate processes like those we have described (a very complete discussion on the determination of the intrinsic efficiency can be found in \cite{Crouthamel}). The goal of the fitting is to test the adequacy of the angular distribution (\ref{coinc}) that we obtained.

Having $\beta$ and $A \, \epsilon_1 \, \epsilon_2$ and using (\ref{coinc}), one can actually predict the number of coincidences for a given angle $\alpha$. The results of this article therefore allow for a much more complete and detailed study of the $\gamma-\gamma$ coincidences from a $\Na$ source. Up to now, essentially only a qualitative analysis could be made. The best approximation to our result that one could often find was to consider an intersection of two circles instead of two spherical caps and compute the respective area. That choice, however, is valid when the rays are parallel, i.e. when the source of radiation is at infinity; the corresponding calculation is often considered for lasers and telescopes \cite{Me}. It does not correspond to our case: a point source, whose emitted rays are not parallel.

Throughout this article we always considered a point-like source; a case we did not consider was that of a source with finite size. This study is presented in \cite{Knoll:2000fj} for the counting rate of a single detector: the result is given in terms of Bessel functions. Extending such study to the rate of coincidences we have studied could be the topic of a future project. But, in general, considering a point-like source is a suitable approximation for this experiment.

Another project that one can consider is to extend the computation of the geometrical correction factors to a general angular correlation function, and not just to the two simpler cases of uncorrelated and fully correlated gamma rays we have considered in this article. These factors are left as parameters that can be fitted to experimental data \cite{cruz}, but we believe they may also be obtained analytically. We leave this study to a future work.

%%%%%%%%%%%%%%%%%%%%%%%%%%%%%%%%%%%%%%%%%%%%%%%%%%%%%%%%%%%%%%%%
\section*{Acknowledgments}
%%%%%%%%%%%%%%%%%%%%%%%%%%%%%%%%%%%%%%%%%%%%%%%%%%%%%%%%%%%%%%%%
I wish to thank Paula Bordalo and S\'ergio Ramos, my professors of Experimental Nuclear Physics at Instituto Superior T\'ecnico in Lisbon: in this undergraduate course, I had the opportunity to perform the experiment of measuring the angular dependence of the $\gamma-\gamma$ coincidences from a $\Na$ source and become familiar with this subject. I also thank my lab partners Armando Fernandes and Pedro Castelo Ferreira, and the referees for pertinent suggestions. I also wish to acknowledge useful discussions with Gustavo Granja and the hospitality of the Center of Mathematics of University of Minho, where parts of this work were completed. This work has been supported by Funda\c c\~ao para a Ci\^encia e a Tecnologia under contract IT (UID/EEA/50008/2019).

%%%%%%%%%%%%%%%%%%%%%%%%%%%%%%%%%%%%%%%%%%%%%%%%%%%%%%%%%%%%%%%%%
%%%%%%%%%%%%%%%%%%%%%%%%%%%%%%%%%%%%%%%%%%%%%%%%%%%%%%%%%%%%%%%%%

\end{document}